\documentstyle[multicol,aps,prl,psfig]{revtex}
\preprint{HLRZ 94--18}
\def\mif{{\hspace{0.5cm} \rm if} \hspace{0.5cm}}
\def\mwith{{\hspace{0.5cm} \rm with} \hspace{0.5cm}}
\def\mor{{\hspace{0.5cm} \rm or} \hspace{0.5cm}}
\def\mconst{{\rm const.}}
\begin{document}
\title{{\small Int. J. Bifurcation and Chaos 5, 51--61 (1995)}\\[0.5cm]
Entropy, Transinformation and Word Distribution \\ of 
Information--Carrying Sequences}
\author{Werner Ebeling\cite{bylinewe}, Thorsten P\"oschel\cite{bylinetp}}
\address{Institute for Theoretical Physics, Humboldt--Universit\"at zu
Berlin, \\Invalidenstra\ss e 42, D--10099 Berlin}
\author{Karl--Friedrich Albrecht}
\address{Research Center J\"ulich, D--52425 J\" ulich, Germany}
\date{June 13, 1994}
\maketitle
\begin{abstract}
We investigate correlations in information carriers, e.g. texts and
pieces of music, which are represented by strings of letters.  For
information carrying strings generated by one source (i.e. a novel or
a piece of music) we find correlations on many length scales. The word
distribution, the higher order entropies and the transinformation are
calculated. The analogy to strings generated through symbolic dynamics
by nonlinear systems in critical states is discussed.
\end{abstract}
\setcounter{topnumber}{3}
\renewcommand{\topfraction}{1.0}
\begin{multicols}{2}
\section[intro]{Introduction}
In the manifold of structures which are used as information carriers
in nature, culture and engineering, linear strings consisting of
sequences of letters play a central role. This may be demonstrated by
the following examples: The proteins and polynucleotids, the main
information carriers in living systems are sequences of amino acids
and/or
nucleotides~\cite{gatlin,ebeling_engel_feistel,feistel_ebeling,nicolis}. Further
most of the messages transporting information between human
informational systems have the form of strings of letters. Examples
are books or letters, music, computer programs etc. By using the
methods of symbolic dynamics any trajectory of a dynamic system may be
mapped to a string of letters on certain
alphabet~\cite{nicolis}\cite{ebeling_nicolis_92}. Hence each sequence
can be interpreted as the trajectory of a discrete dynamic system.
\par
This work is devoted to the investigation of strings of the type
characterized above, i.e.~to objects (documents, programs etc.) which
may be mapped to strings of letters. The main aim of the investigation
is the analysis of long range correlations in information carrying
strings. For several reasons we expect the existence of long range
structures in these sequences~\cite{schenkel_93}.  Especially we
expect correlations which range from the beginning of a string up to
its end. Let us discuss some of the reasons for this behavior:
\begin{enumerate}
\item Predictability: We know from our every-day experience and from scientific
research that the identification of the first hundred or thousand
letters of the string tells us already a lot about the
continuation. Often we make a decision in a book shop after reading
just one page. For example, if we find there several times the word
{\it love} and {\it tennis} we expect to find them on the other pages
again and again, but if we find first words like {\it file} and {\it
program} we expect to remain in quite another field. Listening to the
beginning of a Bach praeludium where the general theme is worked out
we expect to hear it again in many variations up to the very end. Such
expectations are only justified if there exist indeed long range
correlations. This is the scientific expression of our intuitive
expectations which are based on the experience that texts and music
have certain inherent predictability.
\item Syntactical limitations: Another heuristic reason to expect long range 
correlations is the exponential explosion of the variety of possible
subwords with increasing length for uncorrelated strings. Uncorrelated
sequences generated on an alphabet of $\lambda$ letters have a variety
of
\begin{equation}
N(n) = \lambda^n=\exp\left( \ln \lambda \cdot n \right)
\label{manifold}
\end{equation}
different subwords of length $n$. A subword is here any combination of
letters including the space, punctuation marks etc. Strings of this
type may be generated by stochastic processes of Bernoulli--type or by
fully developed chaotic dynamics using alphabets of $\lambda$
letters. For $n >100$ the number $N(\lambda)$ is extremely large. In
other words we need very sharp restrictions to select a meaningful
subset out of it. Long range correlations provide such a selection
criterion. Denoting the selected subset by $N^*(n)$ we expect
\begin{equation}
\frac{N^{*}(n)}{N(n)} \longrightarrow 0 \mif n\longrightarrow\infty
\end{equation}
Bounds of this kind are given by syntax and semantics. The syntactical
rules do not allow for an arbitrary concatenation of words to
sentences, most of them are forbidden.  Furthermore we know that texts
(and pieces of music) are formed by keywords (motifs) which are the
raw material for the generation of a text (a piece of music). In fact
all these rules lead to slower growth with $n$.
\par
A rather sharp restriction on the growth with $n$ corresponds to the
power law
\begin{equation}
N^{*}(n)\sim n^{\alpha}~.
\label{scaling}
\end{equation}
Symbolic strings generated by the logistic map in the
Feigenbaum--point have this scaling
property~\cite{ebeling_nicolis_92}.  The conjecture that several
natural objects have this type of scaling has been made
earlier~\cite{ebeling_feistel_herzel}. We will show here that pieces
of music possibly belong to this class. Another growth law which is
much faster is the stretched exponential law
\begin{equation}
N^{*}(n)\sim \exp \left( Cn^{\alpha} \right) \mwith \alpha < 1 ~.
\label{scaling_rule}
\end{equation}
This scaling which is typically for intermittent processes was
observed for several texts~\cite{ebeling_nicolis_92}. We mention that
in the limit $\alpha \rightarrow 0$ this law corresponds to the
scaling in eq.~(\ref{scaling}).  The reduction due to the scaling
rule~(\ref{scaling_rule}) is not as strong as in the scaling
rule~(\ref{scaling}), however, the reduction factor corresponding to
rule~(\ref{scaling_rule}) is still enormous for large $n$.
\item Evolution: The third reason to expect such behavior is the
  rather general idea that evolution operates in regions where long
  relaxation times, $1/f$-noise and other long range correlations are
  essential~\cite{bak_chen_creutz,kauffman,langton,anishchenko_ebeling_neiman}. 
\end{enumerate}
\section[measure]{Entropy--like Measures of Sequence Structure}
In the following section we will study entropy-like quantities as a
measure of the long range correlations in sequences. In order to
describe the structure of a given string of length $L$ using an
alphabet of $\lambda$ letters $\{A_1 A_2 \dots A_\lambda \}$ we
introduce the following notations~\cite{ebeling_nicolis_91}:
\par \noindent
Let $A_1 A_2 \dots A_n$ be the letters of a given substring of length
$n\le L$. Furthermore let
\begin{equation}
p^{(n)}(A_1\dots A_n)
\end{equation}
be the probability to find in a string a block of length $n$ (subword
of length $n$) with the letters $A_1\dots A_n$. The probability of
having a pair with $(n-2)$ arbitrary letters in between we denote by
\begin{equation}
p^{(n)}(A_1,A_n)=p^{(n)}(A_1 {\rm ~?~?~?~?~ } A_n) ~.
\end{equation}
We introduce the following quantities:
\begin{enumerate}
\item the mutual information, also called transinformation~\cite{li}
\begin{equation}
I(n)=\sum_{A_i A_j} p^{(n)}(A_i,A_j)\log 
\left( \frac{p^{(n)}(A_i,A_j)}{p^{(1)}(A_i)\cdot p^{(1)}(A_j)}\right) ~,
\label{transinformation}
\end{equation}
\item the entropy per subword of length $n$
\begin{equation}
H_n=-\sum p^{(n)}(A_1 \dots A_n) \log p^{(n)}(A_1 \dots A_n)~,
\end{equation}
\item the uncertainty of the letter following a block of length $n$
\begin{equation}
h_n = H_{n+1}-H_n ~,
\end{equation}
\item the  entropy  of  the  source  (related to the 
Kolmogorov--Sinai entropy)
\begin{equation}
h=\lim_{n \to \infty} h_n ~.
\label{uncernity}
\end{equation}
\end{enumerate}
In an earlier paper \cite{ebeling_nicolis_91} we 
assumed the scaling behavior
\begin{equation}
\begin{array}{l}
H_n=n\cdot h + g\cdot n^{\mu_0}\cdot(\log n)^{\mu_1} + e \\
0\le \mu_0 <1 \mor \mu_0 =1 ~~,~~ \mu_1 <0 ~.
\end{array}
\label{ebeling_scaling}
\end{equation}
Related assumptions have been made independently by several
authors~\cite{grassberger_90,li,hilberg_90,schenkel_93}.  From that
point of view the long range order of strings may be well
characterized by the asymptotic for large $n$. Chaotic and stochastic
strings of the standard type have the property $h>0$. Special cases
are Bernoulli processes or fully developed chaos with
\begin{equation}
\begin{array}{l}
p^{(n)}(A_1\dots A_n)=a/\lambda ^n \\
H_n=n\cdot \log \lambda \\
h_n=\log \lambda ~.
\end{array}
\end{equation}
In the following we write all entropy measures in units of
$\log\lambda$.  For Markov processes with memory $m$ the uncertainty
decreases during the first $m$ steps and remains then constant
\begin{equation}
h_{m+k}=h ~~,~~ k>0 ~.
\end{equation}
On the other hand any string with a finite period $p$ is characterized
by
\begin{equation}
H_{p+k}=\mconst ~~,~~ h_{p+k}=0 \mif k>0 ~.
\end{equation}
This means that any new letter added to a string does not increase the
complexity of the sequence. Consequently we find for periodic strings
$h=0$ and $g=0$.  Of special interest to our further considerations
are systems which are neither periodic nor chaotic and which are
characterized by
\begin{equation}
g>0 ~~,~~ h\ll 1 ~.
\end{equation}
Presently there are not enough data available to estimate the limit
entropy $h$ for homogeneous texts (written by one author). We follow
in this respect the seminal work by Claude E. Shannon who concluded in
his pioneering paper: ``From this analysis it appears
that, in ordinary literary English, the long range statistical effects
(up to 100 letters) reduce the entropy\dots''. Shannon gave an
estimate of 40 bits for the entropy of $n=100$ letters, i.e. about 0.4
bit per letter. Transforming the bits to $\lambda$--units (which we
use throughout this paper) we get $H_{100}=8$ and we find for the
entropy per letter the value $0.08$. According to several general
investigations~\cite{levitin_reingold} it is not likely that the
uncertainty (entropy per letter) decreases still further for larger
$n$n. Based on these considerations we assume in the following that
the limit entropy (in $\lambda$--units) is in the region
\begin{equation}
0.01 \le h \le 0.1 ~.
\end{equation}
Since a reliable estimate seems to be impossible at present we simply
neglect the contribution $nh$ to the block entropy in
eq.~(\ref{ebeling_scaling}). In this way we obtain for the
intermediate region $1\ll n \ll 30$ the formula which will be the
basis for our further investigations
\begin{equation}
H_n = g\cdot n^{\mu_0}\cdot (\ln n)^{\mu_1}+e ~.
\end{equation}
Special cases are: 
\begin{itemize}
\item logarithmic tails
\begin{equation}
H_n=g\cdot (\ln n)^{\mu_1}+e ~~,~~ \mu_1<0
\label{logtails}
\end{equation}
\item power law tails
\begin{equation}
H_n=g\cdot n^{\mu_0}+e ~~,~~ 0<\mu_0<1
\label{powertails}
\end{equation}
\end{itemize}
The working hypothesis developed earlier~\cite{ebeling_nicolis_91} is
that strings characterized by eqs.~(\ref{logtails}) or
(\ref{powertails}) being on the borderline between order and chaos
might be prototypes of information carrying sequences.
\par 
Following a relation derived by McMillan and Khinchin the $n$--letter
entropy and the mean number of subwords of length $n$ are
related~\cite{ebeling_nicolis_92}
\begin{equation}
N_n^* = \lambda ^{H_n} ~.
\label{nstar}
\end{equation} 
In this way a logarithmic entropy scaling corresponds to a power law
of the numbers of subwords and a power law scaling of the entropy
corresponds to a stretched exponential growth of the number of
subwords.
\par
The mutual information (transinformation) defined by
eq.~(\ref{transinformation}) is not monotonically decreasing with
increasing $n$. We define here long range tails as any non exponential
decay or increase of the averaged transinformation $I(n)$.  Periodic
strings show correlations of infinite long range. Periodicity with the
period $p$ implies that all conditional probabilities
$p^{(n)}(A_iA_j)/p^{(1)}(A_i)$ for $n>0$ are also periodic.  This
leads to a periodic behavior of the transinformation. Therefore
sequences with long range correlations show a fluctuating behavior at
large scales~\cite{li,herzel}.
\section{Mutual Information and Word Di\-stri\-but\-ions for Finite 
Information--Carrying Strings} 
Printed texts in natural languages and music written in the language
of notes are examples of information--carrying strings. Other
examples, which we do not consider here, are biosequences, where some
evidence for the existence of long range correlations was
found~\cite{jimenez_montano,voss,peng,schenkel_93}.  Originally texts
or pieces of music were generated by the writer or composer as a
dynamical process in real time. Today we find in books or albums the
frozen in results of this process in the form of a symbolic
sequence. Certainly texts or pieces of music are symbolic sequences of
high complexity. In contrast to other dynamical processes, writing and
composing have developed during a long way of evolution being intended for
communication between human beings. In spite of all these difficulties
let us now follow the largely simplifying assumption due to Shannon,
Mandelbrot and others that texts and pieces of music may be considered
as the results of a stationary random process. Although this
assumption is still controversial we will take it here as a basis for
the further analysis. In our analysis we considered the following
sequences:
\begin{enumerate}
\item Sonata for piano forte op. 31 No 2 by L.~v.~Beethoven ($L\approx 4,920$)
\item Moby Dick by H.~Melville ($L\approx 1,170,200$)
\item Grimm's Fairy Tales by the Brothers Grimm ($L\approx 1,435,800$)
\end{enumerate}
Furthermore a few comparisons were made with the Praeludium in
F--Major by J.~S.~Bach and with the sonata KV 311 by W.~A.~Mozart.  In
the case of texts we used an alphabet consisting of the 32 symbols \\
\par
\noindent
\centerline{\fbox{a b c d e f $\dots$ x y z , . ( ) \# $\sim$}}\vspace{0.3cm}\\
The last symbol $\sim$ stands for the empty space and \# for any
number. In the case of music we encoded the notes for $2\frac{1}{2}$
octaves beginning with the low $A$ and ending with the high $D$ on an
alphabet with 32 symbols. The white keys on the piano forte where
encoded by the 18 symbols \\
\par
\noindent
\centerline{\fbox{A H C D E F G a h c d e f g m o p r}}\vspace{0.3cm}\\
and the black keys beginning with the lower {\it Be} and ending with
the high {\it Cis} were encoded by the 12 symbols \\
\par
\begin{figure}[htbp]
\begin{minipage}{8 cm}
\centerline{\psfig{figure=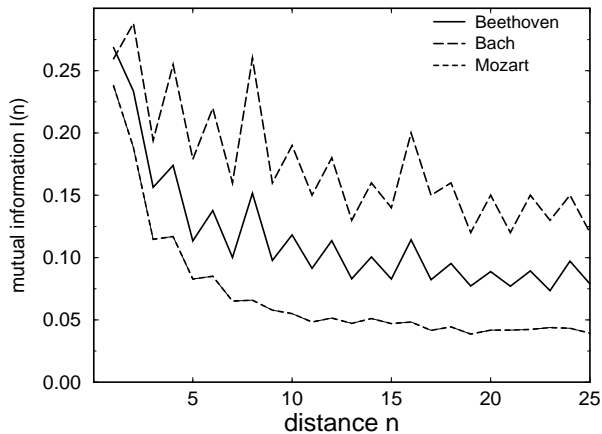,width=8cm,angle=270}}
\caption{The mutual information calculated for Beethoven's Sonata 32
No.~2. For comparison the results for pieces by Bach and Mozart 
from other sources [Ebeling,1993] [Albrecht~et.~al.,1994] are given.}
\label{mutual_beethoven_bach_mozart}
\end{minipage}
\end{figure}

\noindent
\centerline{\fbox{B I J K L b i j k l n q}}\vspace{0.3cm}\\
The pause was encoded by the score ``\_'' and holding a tone by the
``--''.  In order to get a better statistics we also used compressed
alphabets consisting of 3 letters $O$, $M$, $L$ only.  The letter $O$
codes for vowels in the case of texts or for a move down in the case
of music, the letter $M$ stands for consonants or move ups, the letter
$L$ codes for all other symbols, e.g.~pauses (spaces), holding the
tone and punctuation marks.  For the analysis of the mutual
information we have to count here the frequencies of pairs. Since the
number of different pairs is $32^2=1024$ we have a rather good
statistics if the length $L$ of the string satisfies the inequality
$L\gg 1024$.  As shown by several
authors~\cite{li,voss,herzel_schmitt_ebeling_csf} the transinformation
is a reliable measure of the correlations of letters at the distance
$n$.  Every peak at $n$ in the transinformation corresponds to a
strong positive correlation.  In
fig.~\ref{mutual_beethoven_bach_mozart} we show the pair correlations
of the Beethoven sonata and of music by Bach and
Mozart~\cite{ebeling_music}\cite{ebeling_froemmel}.  The peaks show
that there exist strong correlations between two notes at
characteristic distances. The interpretation of these peaks must be
left to specialists. We further notice some similarity in the
correlation structure of Bach's and Beethoven's music and a distinct
different structure of Mozart's music. A more detailed study of the
differences between composers will be given in a separate
paper~\cite{ebeling_froemmel}.

\begin{figure}[htbp]
\begin{minipage}{8 cm}
\centerline{\psfig{figure=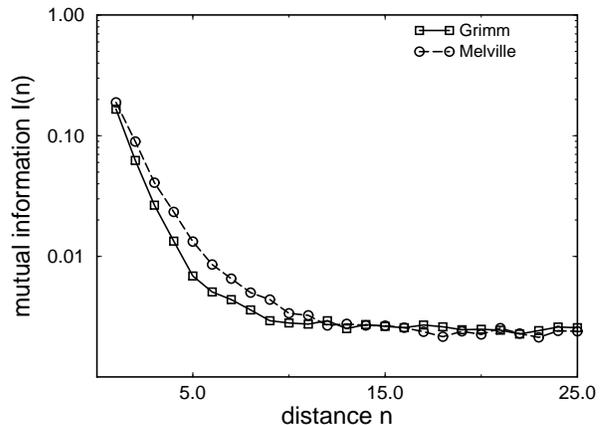,width=7.3cm,angle=270}}
\caption{The mutual information calculated for Moby Dick 
and for Grimm's Tales ($\lambda=32$).}
\label{mutual_moby_grimm}
\end{minipage}
\end{figure}
\begin{figure}[htbp]
\begin{minipage}{8 cm}
\centerline{\psfig{figure=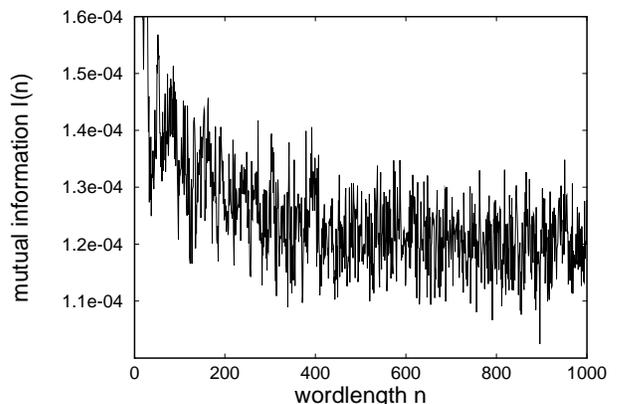,width=8cm,angle=270}}
\caption{The mutual information ($I(n)$, $\lambda =32$) calculated
for the range $n \in [20;1000]$ for Moby Dick. The fluctuation
level due to the finite length of the text ($L=1,170,200$) is 0.00012~.}
\label{trans_moby}
\end{minipage}
\end{figure}
In Figs.~\ref{mutual_moby_grimm} and~\ref{trans_moby} the mutual
information calculated for Moby Dick and for Grimm's Tales
($\lambda=32$) are drawn. The results show that there are well
expressed correlations in the range $n=1\dots 25$ which are followed
by a long slowly decaying tail. The results for the transinformation
$I(k)$ become meaningless if the values are smaller than the level of
natural fluctuations due to the finite length $L$ of the text which
is~\cite{herzel_schmitt_ebeling_csf}
\begin{equation}
\delta I=\frac{\lambda^2-2\cdot\lambda}{2\cdot L \cdot \ln \lambda }~.
\end{equation}
Although the fluctuation level decays with $1/L$ it has even for the
rather long text Moby Dick with $L=1,170,200$ a value of about
0.00012. This means, as seen from Fig.~\ref{trans_moby} that any
conclusions suggesting long range correlations beyond $n=300$ are
rather problematic. However, the range of studies of $I(k)$ may still
be extended by using length corrections
\cite{herzel_schmitt_ebeling_csf}. 
An alternative
method is based on studies of the dependence of $I(k)$ on $1/L$ (which
presumably is linear) and by extrapolations  $(1/L) \rightarrow
\infty$ \cite{ebeling_nicolis_92,herzel_schmitt_ebeling_csf}.
\par
As we see from Figs.~\ref{mutual_moby_grimm} and~\ref{trans_moby} the
mutual information decreases monotonously up to $n \approx 300$ and
converges into the fluctuation level. There are no well expressed
correlation peaks.  Evidently long texts are in this respect less
correlated than DNA sequences where well expressed long range pair
correlations have been found~\cite{herzel_schmitt_ebeling_csf}.
\par
The results obtained so far base only on the statistical distributions
of pairs.  In this case one can reach a rather good statistics by
counting the probability of pairs.  Let us study now the distribution
of words of length $n>2$.  Due to the fact that the number of
different words of length $n$ using an alphabet consisting of 32
letters is
\begin{equation}
N^{*}_n \sim 32^n=2^{5\cdot n}
\end{equation}
there are for $n=9$ already $2^{45}$ words with approximately $2^{50}$
letters.  This is much more than all texts stored in libraries which
have been estimated to consist of about $10^{15}\approx 2^{18}$
letters~\cite{levitin_reingold}.  Therefore we could not do a
statistic analysis if there were no additional constraints due to
grammar and semantics. According to an earlier
estimate~\cite{ebeling_nicolis_92} we expect that the growth law
scales as
\begin{equation}
N^*_n\sim \exp\left(A\cdot n^\alpha\right)
\end{equation} 
with $\alpha \approx \frac{1}{2}$
for texts.
\par
\begin{figure}
\begin{minipage}{8 cm}
\centerline{\psfig{figure=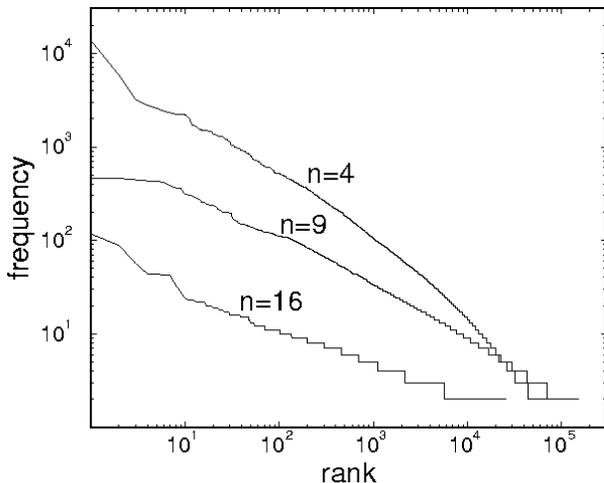,width=8cm}}
\caption{The observed rank--ordered distribution of words of length  
$n=4, 9, 16$ for Moby Dick.}
\label{rank_moby}
\end{minipage}
\end{figure}
\begin{figure}[htbp]
\begin{minipage}{8 cm}
\centerline{\psfig{figure=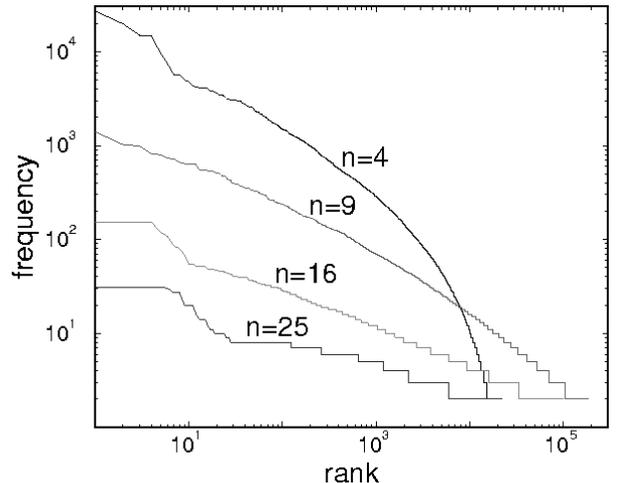,width=8cm}}
\caption{The observed rank--ordered distribution of words of length 
$n=4, 9, 16, 25$ for Grimm's Tales.}
\label{rank_grimm}
\end{minipage}\end{figure}

We have done the analysis with two long texts, namely Moby Dick and
Grimm's Tales. Figs.~\ref{rank_moby} and~\ref{rank_grimm} show the
rank ordered distribution of subwords of length $n=4,~9,~16,~25$.  The
structures of both distributions are similar but the lists of words
are quite different. For example among the most frequent subwords of
length $n=16$ in the case of Moby Dick are ``{\it
the\_sperm\_whale\_}'', and ``{\it the\_quarter\_deck}''. For Grimm's
Tales rather frequent subwords of length $n=25$ are e.g. ``{\it
\_if\_i\_could\_but\_\-shudder.\_}'' and 
``{\it \_princess,\_open\_the\_door\_}''.

The shapes of the subword distributions are distinctly not Zipf--like,
they do not follow a power law. The distribution tends to form a
Fermi--like plateau.  This tendency is based on the theorem of
asymptotic equipartition derived by McMillan~\cite{mcmillan} and
Khinchin~\cite{khinchin}. This theorem proves that for $n\rightarrow
\infty$ the asymptotic form of the distribution is rectangular, i.e.
\begin{equation}
Z(j) = \left\{ \begin{array}{rl}
                        1/N^*(n) & if \quad i<N^*(n)\\
                        0 & else ~.
                \end{array} \right.
\label{rectangular} 
\end{equation}
The effects due to finite $n$ and to the finite text length $L$ tend
to smooth the edges of its distribution. Since our texts are rather
long the finite size effects do not have too much influence to the
distributions.  Much more difficult is the analysis of relative short
pieces of music. Here additional problems arise due to the short
sample. For example Beethoven's sonata consists of only 4,920 letters
(notes). The importance of length corrections for estimating the
frequencies of words was considered by several
authors~\cite{peng}. For a deeper analysis of this problem we refer
to~\cite{herzel_schmitt_ebeling_csf}. The method we used here was
found by generalizing a method proposed in
\cite{schmitt_herzel_ebeling_epl}. According to this method the  
unknown distribution function of the words is guessed by a comparison
of expected (generated) and observed distributions.  Instead of the
simple rectangular distributions in eq.~(\ref{rectangular}) we applied
a more realistic expression.  For given word length $n$ we guess the
true (not normalized) distribution, i.e. the distribution for $L
\rightarrow
\infty$, in the form
\begin{equation}
Z(j,n)=Z_0(j,n)+Z_1(j,n)+Z_2(j,n)
\end{equation}
with
\begin{eqnarray}
Z_0(j,n) & = & \frac{z_0(n)}{1+\exp\left( b\cdot (j-j_0(n))\right)} \\
Z_1(j,n) & = & z_1(n)\cdot \exp \left( -\frac{j-1}{j_1(n)} \right)\\
Z_2(j,n) & = & \left(z_2(n)-z_1(n)-z_0(n)\right)\cdot \nonumber\\
&& \exp \left(-\frac{((j-1)/j_2(n))^2}{2} \right)  
~.
\end{eqnarray}
This distribution has 7 free parameters, one of them is fixed by the
condition that the total number of words is given by the size of the
text. For a string of length $L$ the total number of $n$--words is $N
=L-n+1$. Each of the parameters has a simple meaning as: \\
\begin{tabular}{lcl}
$z_2(n)$ & - & frequency of the most frequent word of length n,\\
$z_1(n)$ & - & height of the exponential contribution at $j=1$,\\
$z_0(n)$ & - & height of the asymptotic ``Fermi''-plateau,\\
$j_2(n)$ & - & number of frequent words,\\
$j_1(n)$ & - & number of relatively frequent words,\\
$j_0(n)$ & - & number of different words,\\
$b(n)$  & - & reciprocal ``Fermi''-temperature of the plateau.
\end{tabular}
\vspace{0.2in}
\par
All these parameters characterize certain properties of the assumed
generating process. For complex processes such as writing of novels or
pieces of music the parameters are of course unknown. Due to their
simple meaning, however, it is easy to guess the initial values of the
parameters. The next step in an iterative process is then to generate
a sample of $n$--words (with the help of the guessed probability
distribution $Z(j,n))$ which has the same number of $n$--words as the
text and transform it to a rank--ordered distribution. To overcome
problems of low statistics we average over several generated
distributions.  The result of this procedure is called the expected
distribution $Z^{exp}(j,n,L)$ according to the chosen set of
parameters. The parameters are fitted by adaptation of the expected to
the empirical (observed) distribution $Z^{obs}(j,n,L)$. The procedure
is iterated up to a sufficient agreement between the observed and the
expected distributions
\begin{equation}
\sum_{j=1}^{N}\left[ Z^{obs}(j,n,L)-Z^{exp}(j,n,L) \right]^2 \rightarrow min ~.
\end{equation}

\begin{figure}[htbp]
\begin{minipage}{8 cm}
\centerline{\psfig{figure=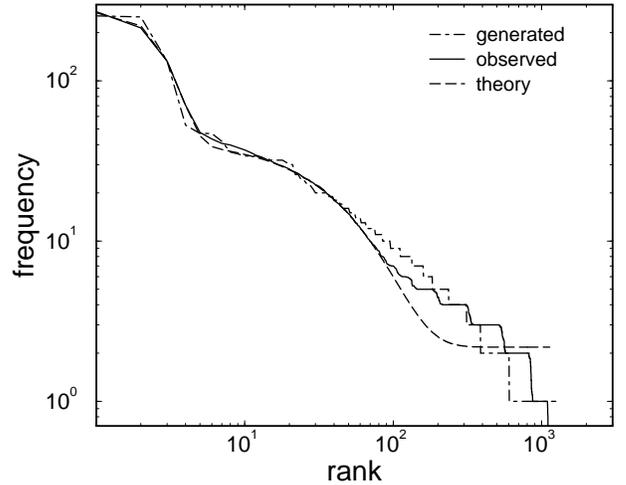,width=8cm,angle=270}}
\caption{Rank ordered word distribution for Beethoven's sonata ($n=8,
\lambda=3$): Observed frequency (full line), generated frequency (dashed line)
and theoretical curve (dotted line).}
\label{distr_beethoven}
\end{minipage}\end{figure}

\begin{figure}[htbp]
\begin{minipage}{8 cm}
\centerline{\psfig{figure=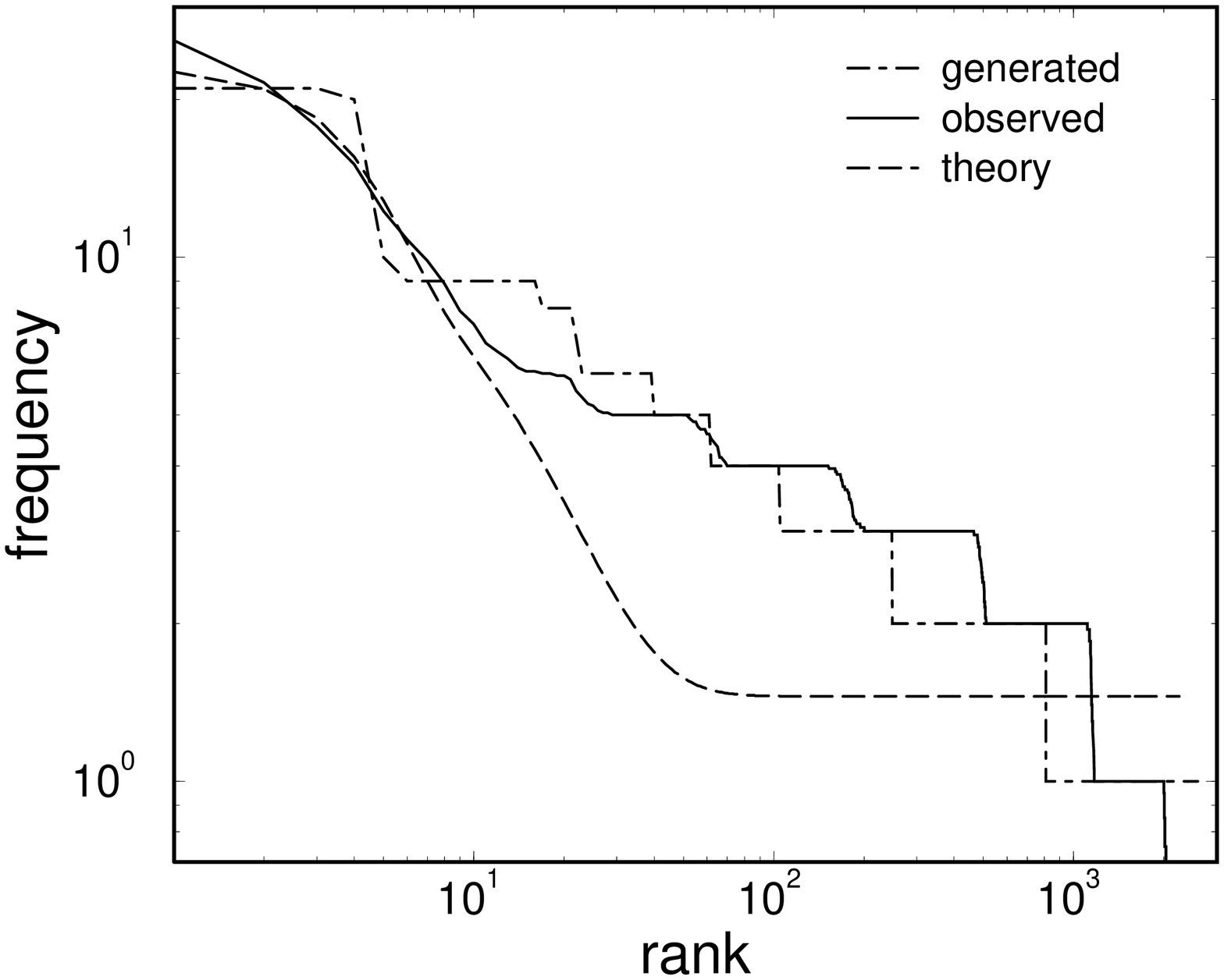,width=8cm,angle=270}}
\caption{Rank ordered distribution for Beethoven's sonata ($n=8,
\lambda=32$): Observed frequency (full line), generated frequency (dashed line) 
and theoretical curve (dotted line).}
\label{beethoven_32}
\end{minipage}\end{figure}

In the minimization procedure we used the {\it simplex}--method
applying the program package {\sc MINUIT}~\cite{minuit} developed for
non--linear parameter optimization. For word length $n=8$ the
optimization parameters are given in Tab.~\ref{tabelle}.  The
corresponding distributions $Z(j,n)$, $Z^{exp}(j,n,L)$ and
$Z^{obs}(j,n,L)$ are presented in figs.~(\ref{distr_beethoven}) and
(\ref{beethoven_32}).

\begin{table}
\begin{tabular}{ccccccccc} 
alphabet&$Z_2(8)$&$Z_1(8)$&$Z_0(8)$&$j_2(8)$&$j_1(8)$&$j_0(8)$&$b(8)$&$F(8)$
\\ \hline
 3 & 241 & 37.5 & 2.54 & 1.44 & 41.4 & 908 & 0.28 & 2386 \\
32 & 19.4 & 9.82 & 1.25 & 2.8 & 10.8 & 3733 & 0.317 & 1381 
\end{tabular}
\caption{Parameters of the distribution of words (sequences of notes) of
length $n=8$ for Beethoven's Sonata.}
\label{tabelle}
\end{table}

\section{Analysis of Entropy Scaling}
As shown above the set of entropies of subwords is the basic measure
for the investigation of the order of a string. The probabilities one
need for the calculation of the entropies are in general unknown and
have to be estimated from the frequencies of words. The theoretical
value of the block entropy follows from the distribution discussed in
the last section. For simplicity we will omit in the following the
block length $n$. The theoretical entropy is then
\begin{equation}
H=-\sum_i \frac{Z(i)}{N} \cdot \log\left(\frac{Z(i)}{N}\right) ~. 
\label{correct}
\end{equation}
The calculation of the block entropy from the distribution function
$Z(i)$ is rather difficult and time consuming. The direct observation
of the entropy is based on the frequencies of the different
words. Replacing the probabilities by the observed frequencies $N_i/N$
in the entropy definition we obtain the observed entropy
\begin{equation}
H^{obs}=\log(N)-\frac{1}{N}\sum_iN_i\cdot\log\left( N_i\right) ~.
\end{equation}
This is a random variable with the expectation value
\begin{equation}
H^{exp}=\langle H^{obs}\rangle=\log(N)-\frac{1}{N}\sum_i\langle
N_i\cdot \log
\left(N_i\right)\rangle ~.
\end{equation}
Let us assume now that the subwords are Bernoulli distributed.  Then
we find (omitting further the index $n$ which characterizes the length
of the subwords) for the mean value
\begin{eqnarray}
\langle H^{obs} \rangle &=& \log N - \sum_i \sum_{N_i} \frac{(N-1)!}{(N_i-1)!~
(N-N_i)!}\cdot p_i^{N_i} \nonumber \\
&&\cdot (1-p_i)^{N-N_i}\cdot\log N_i ~.
\end{eqnarray}
We will show now that the coefficients in this expansion are closely
related to the Renyi entropies of order $q$ which are defined by
\cite{herzel_schmitt_ebeling_csf} 
\begin{equation}
H^{(q)}=\frac{1}{1-q}\cdot\log \left(\sum_i p_i^q \right) ~.
\end{equation}
The Shannon entropies correspond to the limit $q \rightarrow 1$ and
the case $q = 0$ is related to the total number of different subwords
$s$:
\begin{equation}
H = H^{(1)} ~~,~~ s = \lambda^{H^{(0)}} ~.
\end{equation}
In the limit of very long strings, i.e. $N \gg s$, the expected
entropies are given by the
approximation~\cite{herzel_schmitt_ebeling_csf,levitin_reingold}
\begin{equation}
\langle H^{obs} \rangle = H - \frac{s}{2N \log\lambda} 
\label{en_approx_a} ~.
\end{equation}
In the opposite case, where $N \ll s$, the subwords may appear only
very few times. ~\cite{herzel_schmitt_ebeling_csf} Therefore we obtain
the series
\begin{eqnarray}
\langle H^{obs} \rangle &=& \log N - N \cdot \log 2 \cdot \sum_i {p_i}^2
\nonumber \\
&&- \frac{N^2}{2!}\left( \log 3 - 2\cdot\log 2\right)\cdot \sum_i
{p_i}^3 + ...
\label{en_approx_b}
\end{eqnarray}
Using the definition of the Renyi entropies follows
\begin{eqnarray}
\langle H^{obs}\rangle&=&\log N -N\cdot\log 2 \cdot
\lambda^{-H^{(2)}}\nonumber \\
&&-\frac{N^2}{2!} \cdot \left(\log 2-2\cdot\log 2\right) \cdot
\lambda^{-2H^{(3)}} - ...
\label{second_order}
\end{eqnarray}
The higher order Renyi entropies can be found by fitting the
coefficients of $N^k$ in this series to the observed entropies.  Since
the entropies decrease monotonously with their order follows that the
second order entropies ($q=2$) which can be obtained from
eq.(\ref{second_order}) represent a lower boundary of the first order
Shannon entropies ($q=1$).
\par
In this way we have obtained now three procedures to derive the
entropies of human writings with a given length $L$:
\begin{enumerate}
\item Calculation of the usual entropy ($q=1$) from the observed
distribution. Estimation of some finite length corrections using
eq.~(\ref{en_approx_a}). This is the standard procedure applied e.g.
in \cite{herzel_schmitt_ebeling_csf}.
\label{method1}
\item Guessing first the ``true'' distribution $Z(i,n)$ on the basis
of the observed distribution $Z^{obs}(i,n,L)$ and calculating then the
entropies (for any $q$) from it. This method is essentially new. It
was tested here but it still needs further elaboration.
\label{method2}
\item Calculation of the higher order entropies (especially $q=2$)
from the deviations between $\log N$ and the observed entropies. This
method which seems to be new too is of limited accuracy and yields
only rough estimates for the second and higher order entropies.
\label{method3}
\end{enumerate}
Let us consider now our examples: Beethoven's Sonata is rather short
($L=4920$). Therefore the first approach is restricted to $n<7$ for
the 3--symbol alphabet and to $n<3$ for the 32--symbol alphabet. We
have obtained here as a new result the entropy for $n=8$ using method
(\ref{method2}).  The result is
\begin{eqnarray}
H_8/\log(3)&=6.39\hspace{1cm}& {\mbox{\it \footnotesize 3 symbols}} \\
H_8/\log(32)&=2.49\hspace{1cm}& {\mbox{\it \footnotesize 32
symbols}}~.
\end{eqnarray}
For $9\le n \le 26$ the second order entropies were estimated by means
of method (\ref{method3}). The result can be 
approximated by the fit formula
\begin{equation}
H_n/\log 3 \approx 2.04\cdot\left(\log n +1\right) ~~,~~ (\lambda=3)~.
\label{fitformula}
\end{equation}
This fit formula reminds the scaling of type eq.~(\ref{logtails}). The
logarithmic function in eq.~(\ref{fitformula}) yields a better fit
than the power law postulated in an earlier
paper~\cite{ebeling_nicolis_92}.

\begin{figure}[htbp]
\begin{minipage}{8 cm}
\centerline{\psfig{figure=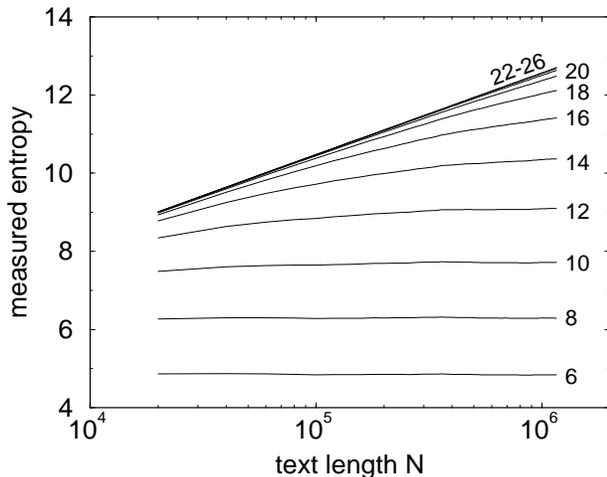,width=8cm,angle=270}} 
\caption{The measured entropy values for Moby Dick over
the length of the text investigated for different word
length $n$.}
\label{en_moby}
\end{minipage}\end{figure}

Due to the rather large length of the text Moby Dick we may apply the
method (\ref{method1}) up to much larger $n$--values. Taking into
account that not all of the combinatorial possibilities are admitted
we went up to $n=16$ for the 3--letter alphabet and up to $n=10$ for
the 32--letter alphabet. The results are given in
table~\ref{tab_e_moby}. We have checked these values also with a
consistency test with method~(\ref{method2}). A more detailed
calculation by means of the distribution function
method~(\ref{method2}) will be discussed in~\cite{poeschel}.  For
larger $n$--values the second order entropies ($q=2$) were estimated
from the differences between $\log N$ and the observed entropies
(Fig.~\ref{en_moby}) using eq.~(\ref{second_order}). The result of our
estimate is given by the fit formulae
\begin{eqnarray}
H_n/\log(3)&=4.8\cdot\sqrt{n}-7.6\hspace{1cm}& (\lambda=3)\\
H_n/\log(32)&=0.9\cdot\sqrt{n}+1.7\hspace{1cm}& (\lambda=32)~. 
\end{eqnarray}
These fit formulae remind a scaling law according to
eq.~(\ref{powertails}) with the exponent 0.5. The low accuracy of the
data does not allow for a quantitative fit of the exponent, in fact
any value between 0.4 and 0.6 gives a reasonable fit. A scaling law of
square root type was first found by Hilberg~\cite{hilberg_90} who
fitted Shannon's original data. This result was reproduced also for a
textbook on
selforganization~\cite{ebeling_engel_herzel,ebeling_nicolis_92}.

Let us consider now the growth of the number of different subwords.
According to relation (\ref{nstar}) the number of different subwords
in Moby Dick grows with $n$ according to a stretched exponential law
\begin{equation}
N_n^*\approx 4.1~10^{-4} \cdot \exp \left( 5.2\cdot \sqrt{n} \right)
~.
\end{equation}
For the sonata the number of different subwords increases not as fast
as for the English text, we find here a quadratic growth law
\begin{equation}
N_n^*\approx 7.6\cdot n^2~.
\end{equation}
\section{Conclusions}
The present paper reports on results concerning information carrying
strings such as texts and pieces of music. The results show that block
entropies and the mutual information are appropriate measures of the
correlations and the degree of order in strings. In agreement with
earlier work~\cite{ebeling_nicolis_92} we have confirmed the
hypothesis that strings produced by information processing sources are
neither periodic nor chaotic but somehow in between. In the present
work we have studied several examples (two books and one piece of
music) to illustrate this hypothesis.
\par
We have shown that there is some empirical evidence for short range
and middle range correlations. This is to be seen in the statistics of
pairs of letters and subwords. Another relevant information is
contained in the block entropies. We developed methods how to guess
these unknown quantities from relatively short samples.  For subword
length $n \gg 100$ we did not find any indication for the existence of
pair or word correlations.  Our empirical studies of the pair
correlation function and the spectrum derived from this
quantity~\cite{anishchenko_ebeling_neiman} show a strong sensitivity
on the length of the texts.  In spite of all these difficulties we may
state in conclusion that several distinct differences from chaotic
strings are observed in texts or pieces of music.
\par
The entropy scaling shows that texts and pieces of music resemble the
sequences created by nonlinear dynamic systems near critical points
through symbolic dynamics.  However, more empirical data on long texts
and pieces of music and a more detailed study of the block entropies
of dynamical systems are needed to elaborate this point further. We
state that with the present methods available so far it is not
possible to calculate high order entropies for $n >30$ since there are
no homogeneous texts which are significantly longer than several
million letters.
\begin{table}
\begin{tabular}{ccccccc} 
$H_2/\log \lambda$         &6   &8   &10  &12 &14  & 16 \\ \hline
$\lambda = 3$ &4.85&6.30&7.72&9.10&10.5&11.54\\ \hline
$\lambda = 32$&3.21&3.92&4.45& & & 
\end{tabular}
\caption{The calculated values of the entropies for Moby Dick.}
\label{tab_e_moby}
\end{table}

\acknowledgments

The authors thank H.~Herzel and A.~Schmitt for fruitful discussions,
C.~Fr\"om\-mel for the permission to compare in
Fig.~\ref{mutual_beethoven_bach_mozart} with an unpublished result on
Mozart's music and the {\em Project Gutenberg Etext, Illinois
Benedictine College, Lisle} for providing the 
ASCII--text of ``Moby Dick'' and ``Grimm's Fairy Tales''.

\end{multicols}
\end{document}